\documentstyle[epsfig]{aipproc}

\def\slash#1{{\mathpalette\c@ncel{#1}}} 
\newcommand\beq{\begin{eqnarray}}
\newcommand\eeq{\end{eqnarray}}

\newcommand\la{\langle}
\newcommand\ra{\rangle}

\begin{document}
\title{Hyperon Polarization in Inclusive Hadronic Production
\footnote[2]{Proceedings of the talk presented at SPIN2000,
Osaka, Oct. 16-21, 2000.}}

\author{Y. Kanazawa and Yuji Koike}
\address{Department of Physics, Niigata University,
Ikarashi, Niigata 950-2181, Japan}

\maketitle

\begin{abstract}
A QCD formula for the polarization in the large-$p_{T}$ $\Lambda$ hyperon 
production in the unpolarized nucleon-nucleon collision at large $x_{F}$ 
is derived.
We focus on the mechanism in which 
the chiral-odd spin-independent twist-3 
quark distribution $E_{F}(x,x)$ becomes the source of the 
transversely polarized quarks fragmenting into the polarized $\Lambda$.  
A simple model estimate for that contribution 
shows the possibility that it gives rise to a sizable $\Lambda$ polarization.
\end{abstract}

It is a well known experimental fact that the hyperons produced 
in the unpolarized nucleon-nucleon collisions are polarized transversely 
to the production plane\,\cite{Bunce76,Smith87}.
In this letter we focus on the polarization 
of the $\Lambda$ hyperon production with 
large transverse momentum in $pp$ collision
\beq
N(P) + N'(P') \to \Lambda(l,\vec{S}_{\perp}) + X.
\label{pplam}
\eeq  
Ongoing experiment at RHIC is expected to provide more data
on the polarization.  
The nonzero $\Lambda$ polarization in this process requires
the presence 
of particular quark-gluon correlation (higher twist effect) 
and/or the 
effect of transverse momentum either
in the unpolarized nucleon or the fragmentation function for $\Lambda$.
According to the generalized QCD factorization theorem, 
the polarized cross section for this process
consists of two kinds of twist-3 contributions:
\beq
&{\rm (A)}&\quad E_a(x_1,x_2)\otimes q_b(x')\otimes \delta D_{c\to \Lambda}
(z)\otimes\hat{\sigma}_{ab\to c},
\label{aterm}\\
&{\rm (B)}&\quad
q_a(x)\otimes q_b(x') \otimes  D_{c\to \Lambda}^{(3)}(z_1,z_2)\otimes 
\hat{\sigma}_{ab\to c}'
\label{bterm}.
\eeq  
Here the functions $E_a(x_1,x_2)$ and $ D^{(3)}_{c\to\Lambda}(z_1,z_2)$
are the twist-3 quantities representing, respectively, the
unpolarized distribution in the nucleon
and the fragmentation function for the transversely
 polarized $\Lambda$ hyperon, 
and $a$, $b$ and $c$ stand for the parton's species.
Other functions are twist-2; 
$q_b(x)$ the unpolarized distribution (quark or gluon) and
$\delta D_{c\to \Lambda}(z)$ the transversity fragmentation function 
for $\Lambda$. 
The symbol $\otimes$ denotes convolution. 
$\hat{\sigma}_{ab\to c}$ {\it etc} represents the partonic cross section
for the process $a+b \to c + anything$ which yields large transverse 
momentum of the parton $c$. 
Note that (A) contains two chiral-odd functions
$E_a$ and $\delta D_{c\to \Lambda}$,
while (B) contains only chiral-even functions. 

In this report, we derive a QCD formula for the polarized cross section 
(\ref{pplam}) from the (A) term in the kinematic region $|x_{F}|\to 1$,
using the valence quark-soft gluon approximation proposed by Qiu and 
Stermann\cite{QS99}.
Employing this approximation, 
they reproduced
the E704 data for the single-transverse spin 
asymmetries in the pion production at $x_F \to 1$ 
reasonably well.
The fact that the perturbative QCD description
for the pion production is valid as low as $l_T\sim$ 1 GeV
encouraged us to
apply the method
to the polarized $\Lambda$ hyperon production (\ref{pplam})
for which the data exist only in the relatively small $l_T$ region.  
At large $x_F>0$, 
which mainly probes large $x$ and small $x'$ region,
the cross section is dominated by the particular terms in (A)
which contain the derivatives of the {\it valence} twist-3 distribution
$E_{Fa}(x,x)$.  The reason for this observation is 
the relation 
$|{\partial \over \partial x}E_{Fa}(x,x)| \gg E_{Fa}(x,x)$
owing to the behavior of $E_{Fa}(x,x)
\sim (1-x)^\beta$ ($\beta >0$) at $x\to 1$.
We thus keep only the terms with the derivative of $E_{Fa}$ for the valence 
quark ({\it valence quark-soft gluon approximation}). 

The polarized cross section for (\ref{pplam}) is a function of 
three independent variables, 
$S=(P+P')^2\simeq 2P\cdot P'$, 
$x_F = 2l_{\parallel}/ \sqrt{S}$ ($=(T-U)/S$), 
and $x_T = 2l_{T}/ \sqrt{S}$.    
$T=(P-l)^2\simeq -2P\cdot l$ and 
$U=(P'-l)^2\simeq -2P'\cdot l$ are given in terms of these three
variables by 
$T= -S\left[ \sqrt{x_F^2 + x_T^2} - x_F\right]/2$ and 
$U= -S\left[ \sqrt{x_F^2 + x_T^2} + x_F\right]/2$.
In this convention, production of $\Lambda$
in the forward hemisphere in the direction of the incident nucleon
($N(P)$) corresponds to $x_F>0$.  
Since $-1<x_F <1$, $0<x_T<1$ and $\sqrt{x_F^2 + x_T^2} < 1$,
$x_F \to 1$ corresponds to the region with $-U\sim S$ and $T\sim 0$.

In the valence quark-soft gluon approximation, the cross section 
for the (A) term reads,
\beq
E_l{d^3\Delta\sigma^A(S_{\perp}) \over dl^3}
&=&{\pi M\alpha_s^2 \over S}\sum_{a,c}\int_{z_{min}}^1
{d\,z\over z^3}{\delta D}_{c\to\Lambda}(z)
\int_{x_{min}}^1 {d\,x\over x}
{1\over xS + U/z}
\nonumber\\
& \times & 
\int_0^1 {d\,x'\over x'}\delta\left(x'+{xT/z \over xS + U/z}\right)
\varepsilon_{l S_{\perp} p n}
\left({1\over -\hat{u}}\right)
\left[ -x {\partial \over \partial x}E_{Fa}(x,x)\right]
\nonumber\\
& \times &
\left[ 
G(x')\delta \widehat{\sigma}_{ag\to c} +\sum_{b} q_b(x') 
\delta \widehat{\sigma}_{ab\to c}
\right],
\label{cross}
\eeq
where 
$p$ and $n$ are the two light-like vectors defined from the momentum of the
unpolarized nucleon as $P=p+M^2n/2$, $p\cdot n=1$ and
$\varepsilon_{l S_{\perp} p n}=\varepsilon_{\mu\nu\lambda\sigma}l^\mu 
S_{\perp}^{\nu} p^\lambda n^\sigma\sim {\rm sin}\phi$ with $\phi$ 
the azimuthal angle between the spin vector of the $\Lambda$ hyperon and 
the production plane.  
The invariants in the parton level are defined as
$\hat{s} =(p_a + p_b)^2 \simeq ( xP + x'P')^2 \simeq xx'S,$
$\hat{t} =(p_a-p_c)^2 \simeq (xP-l/z)^2 \simeq xT/z,$
$\hat{u} =(p_b-p_c)^2 \simeq (x'P'-l/z)^2 \simeq x'U/z$.   
The lower limits for the integration variables are 
$z_{min} = {-(T+U) \over S}=\sqrt{x_F^2+x_T^2}$ and 
$x_{min} = {-U/z \over S+T/z}$.  
$q_b(x')$ is the unpolarized quark distribution, 
and $G(x')$ is the unpolarized gluon distribution.
$\delta \widehat{\sigma}_{ag\to c}$ and $\delta \widehat{\sigma}_{ab\to c}$ 
are partonic cross sections for the quark-gluon and quark-quark processes,  
respectively. 
$E_F(x,x)$ is the soft gluon component of the
unpolarized twist-3 distribution defined as
\beq
E_{Fa}(x,x)={-i\over 2M}
\int{d\lambda \over 2\pi}e^{i\lambda x}\la P | \bar{\psi}^a(0)\!\not{\! n}
\gamma_{\perp\sigma}
\left\{ \int {d\mu\over 2\pi}g F^{\sigma\beta}(\mu n)n_\beta\right\}
\psi^a(\lambda n)|P\ra.  
\label{EF}
\eeq
The summation for the flavor indices of $E_{Fa}(x,x)$ is to be over 
$u$- and $d$- valence quarks,   
while that for the twist-2 distributions is 
over $u$, $d$, $\bar{u}$, $\bar{d}$,
$s$, $\bar{s}$. 
$\delta\widehat{\sigma}_{ab\to c}$ and  
$\delta\widehat{\sigma}_{ag\to c}$ 
can be obtained from the $2\to 2$ cut diagrams.
The result reads 
\beq
\delta\hat{\sigma}_{q q'\to q}
&=&
\left({\hat{s}\hat{u}\over \hat{t}^2}\right)
\left[{2\over 9}+{1\over 9}\left(1+{\hat{u}\over \hat{t}}\right)\right],
\qquad 
\delta\hat{\sigma}_{q \bar{q}'\to q}
=
\left({\hat{s}\hat{u}\over \hat{t}^2}\right)
\left[{7\over 9}+{1\over 9}\left(1+{\hat{u}\over \hat{t}}\right)\right],
\nonumber\\
\delta\hat{\sigma}_{q q\to q}
&=&
-\left({\hat{s}\over \hat{t}}\right)
\left[{10\over 27}+{1\over 27}\left(1+{\hat{u}\over \hat{t}}\right)\right],
\label{qqhard}
\eeq
for $\delta\widehat{\sigma}_{ab\to c}$, and 
\beq
\delta\hat{\sigma}_{a g\to c}
&=&
{9\over 8}\left({\hat{s}\hat{u} \over \hat{t}^2}\right)
+{9\over 8}\left({\hat{u} \over \hat{t}}\right)+{1\over 8}
+\left[{1\over 4}\left({\hat{s}\hat{u}\over \hat{t}^2}\right)
+{1\over 72}\right]
\left(1+{\hat{u}\over\hat{t}}\right).
\label{qghard}
\eeq

We now present a simple estimate of the $\Lambda$ polarization. 
To this end we use a model for $E_F(x,x)$ introduced in Ref. \cite{KK00}. 
It is based on 
the comparison of the explicit form (\ref{EF})
with the transversity distribution
\beq
\delta q_a(x)&=& {i\over 2}\varepsilon_{S_\perp \sigma p n}
\int{d\lambda \over 2\pi}e^{i\lambda x}\la PS |\bar{\psi}^a(0)\!\not{\! n}
\gamma_\perp^\sigma \psi^a(\lambda n)|PS\ra,
\label{transversity}
\eeq
where $\varepsilon_{S_\perp \sigma p n}\equiv \varepsilon_{\mu\sigma\nu\lambda}
S_\perp^\mu p^\nu n^\lambda$.
We make an ansatz 
\beq
E_{Fa}(x,x) = K_a \delta q_a(x),
\label{EFq}
\eeq
with a flavor-dependent parameter $K_a$ which simulates
the effect of the gluon field with zero momentum in $E_F(x,x)$.
We note that even though $E_F(x,x)$ is an unpolarized distribution,
the quarks in $E_F(x,x)$ is ``transversely polarized'' which
eventually fragments into the transversely polarized $\Lambda$. 
The relation (\ref{EFq}) is in parallel with the ansatz originally introduced
in \cite{QS99}
\beq
G_{Fa}(x,x) = K_a q_a(x).  
\label{QSassumption}
\eeq
Here we assume the same parameter for 
$K_a$ in 
(\ref{EFq}) and (\ref{QSassumption})\,\cite{KK00}.  
Since we are only interested in the estimate of
order of magnitude, we do not pay much attention to the 
scale dependence of 
each distribution and fragmentation function
and use those functions
at the scale $1\sim 2$ GeV which is a typical
size of the transverse momentum of the produced $\Lambda$ and pion.
For the unpolarized distribution $q_a(x)$ and $G(x)$,
we use the GRV LO distribution at the scale $\mu=1.1$ GeV
($=l_T$ of $\Lambda$ in R608 data below)\cite{GRV95}. 
For the transversity distribution $\delta q_a(x)$,
we use the GRSV helicity distribution $\Delta q_a(x)$
(LO, standard scenario)\cite{GRSV96} assuming 
$\delta q_a(x)=\Delta q_a(x)$ at the scale $\mu=1.1$ GeV. 
For the fragmentation function of the unpolarized and    
transversely polarized $\Lambda$,
we use, respectively, the fragmentation function of the 
unpolarized and 
longitudinally polarized $\Lambda$, $D_{c\to\Lambda}(z)$ and 
$\Delta D_{c\to\Lambda}(z)$ (three scenarios for $\Delta D$), 
given by de Florian {\it et al}.\cite{FSV98} 
with the assumption that $\delta D_{c\to\Lambda}(z)
=\Delta D_{c\to\Lambda}(z)$ at the scale $\mu=1.1$ GeV.
\begin{figure}[t]
\centerline{\epsfig{file=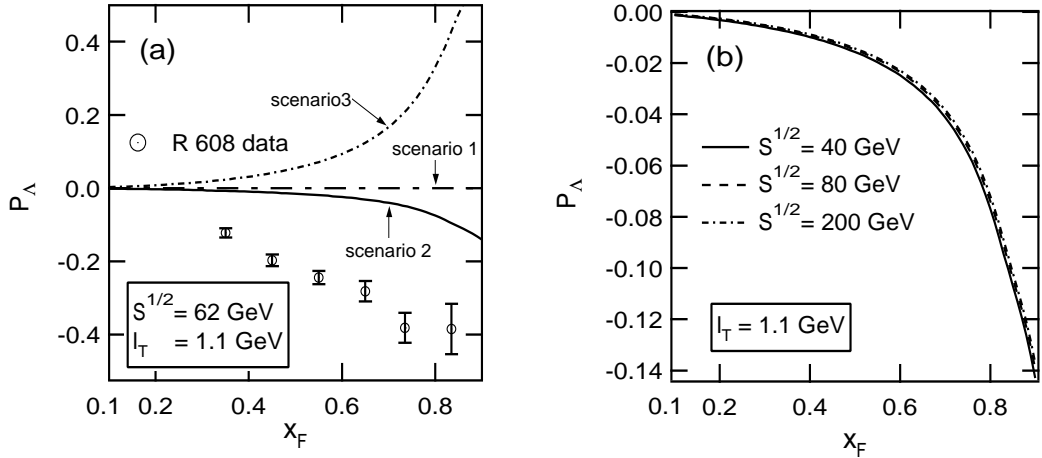,height=7cm,width=15cm}}
\caption
{(a): The R608 data for the polarization $P_{\Lambda}$ 
of the produced $\Lambda$ hyperon 
in the unpolarized proton-proton collision 
at $\sqrt{S}=62$ GeV [2].
The transverse momentum of the $\Lambda$ is 
$l_T=1.1$ GeV. 
The curves show the calculated polarization with three scenarios 
for $\delta D_{c\to \Lambda}$. 
(b): The $\Lambda$ polarization 
in the unpolarized proton-proton collision at 
$\sqrt{S}=40, 80, 200$ GeV
with $l_T=1.1$ GeV. 
Scenario 2 is used for $\delta D_{c\to\Lambda}$. 
}
\label{fig1}
\end{figure}
Following our recent paper\cite{KK00},
we determine $K_{u,d}$ to fit the FNAL E704 data of the single-transverse spin 
asymmetry in the pion production\cite{Adams91} 
using $G_{Fa}(x,x)$ with (\ref{QSassumption}) at the scale $\mu=1.5$ GeV and 
the fragmentation function of the pion given at $\mu=$ 2 GeV in \cite{BKK95}.  
The result is $K_{u}=-K_{d}=0.06$.
The obtained $\Lambda$ polarization 
is shown in Fig. \ref{fig1}(a) for the three scenarios of 
$\delta D_{c\to\Lambda}(z)$ in \cite{FSV98}
with the CERN R608 data\cite{Smith87}.
The scenario 1 corresponds to the expectation from the naive 
non-relativistic 
quark model, where only strange quarks can fragment into a 
polarized $\Lambda$. 
In our approximation, $E_F(x,x)=0$ for the $s$-quark and thus 
the polarization is zero in this scenario. 
The scenario 2 is based on the assumption that the flavor-dependence of 
$\Delta D_{c\to\Lambda}(z)$ is the same as that of  
the polarized quark distribtuions in $\Lambda$
obtained by the SU(3) symmetry
as proposed in Ref. \cite{BJ93}; 
$\delta D_{u\to\Lambda}=\delta D_{d\to\Lambda}=-0.2\delta D_{s\to\Lambda}$. 
In the scenario 3, three flavors of quarks equally fragment into
the polarized $\Lambda$; 
$\delta D_{u\to\Lambda}=\delta D_{d\to\Lambda}=\delta D_{s\to\Lambda}$. 
From this figure, one sees that both scenario 2 and 3 give
rise to increasing polarization at large $x_F$ as expected, 
the former being slightly more favored because of the
sign of the polarization.  
For a complete understanding on the hyperon polarization,
combined analysis of both (A) and (B) contributions 
together with more sophisticated parametrization of
the participating distributions and fragementation functions
is certainly necessary.  

In Fig. \ref{fig1}(b), we plotted the polarization from the term (A) 
for various values of $\sqrt{S}$ at $l_{T}=1.1$ GeV with scenario 2
for $\delta D_{c\to \Lambda}$. 
One sees that the 
result is almost independent of the value of $\sqrt{S}$ in this kinematic 
region.  This tendency is the same as the experimental data.  

A different approach to the $\Lambda$ polarization
introduces
the so-called T-odd distribution or 
fragmentation functions with the intrinsic
transverse momentum instead of 
twist-3 distributions 
introduced here.
Similarly to (A) and (B),
this approach starts from the factorization assumption for the two
types of contributions to the polarization;
(i) $h_{1}^\perp (x,{\bf p}_\perp)\otimes q(x')\otimes \delta D(z)\otimes
\hat{\sigma}$,
(ii) $q(x)\otimes q(x')\otimes D_{1T}^{\perp} (z,{\bf k}_{\perp})\otimes
\hat{\sigma}'$,
where 
$h_1^\perp$ represents distribution of a transversely polarized quark 
with nonzero tranverse mometum inside the unpolarized nucleon, 
and $D_{1T}^{\perp}$ represents a fragmentation function for 
an unpolarized quark fragmenting into a transversely polarized $\Lambda$ 
with the transverse momentum (``polarizing fragmentation function''). 
Anselmino {\it et al.} fitted the experimental data for the $\Lambda$ 
polarization assuming
the above (ii) is the sole origin of the polarization\cite{ABAM00}.  
We expect from the present study, however, that the
large portion of the $\Lambda$ polarization should be ascribed 
to the twist-3 distribution in the unpolarized nucleon
and $\delta D(z)$ which should be related to 
the above contribution (i).
It is interesting to explore the connection between 
the present approach and that in \cite{ABAM00}.  

To summarize, 
we have derived a cross section formula for the polarized $\Lambda$ 
production in the unpolarized nucleon-nucleon collision at large $x_{F}$.  
A simple model estimate for this contribution suggests
a possibility that the contribution from the soft gluon pole 
gives sizable $\Lambda$ polarization.

\end{document}